# Real-Time Dynamic Layout Optimization for Floating Offshore Wind Farm Control

Timothé Jard [1], Reda Snaiki [2*]

[1] Department of Mechanical Engineering, École de Technologie Supérieure, Université du Québec, Montréal, QC, H3C 1K3, Canada

[2] Department of Construction Engineering, École de Technologie Supérieure, Université du Québec, Montréal, QC, H3C 1K3, Canada

*Corresponding author:* reda.snaiki@etsmtl.ca

**Abstract:** Downstream wind turbines operating behind upstream turbines face significant performance challenges due to reduced wind speeds and increased turbulence. This leads to decreased wind energy production and higher dynamic loads on downwind turbines. Consequently, real-time monitoring and control have become crucial for improving wind farm performance. One promising solution involves optimizing wind farm layouts in real-time, taking advantage of the added flexibility offered by floating offshore wind turbines (FOWTs). This study explores a dynamic layout optimization strategy to minimize wake effects in wind farms while meeting power requirements. Two scenarios are considered: power maximization and power set-point tracking. The methodology involves a centralized wind farm controller optimizing the layout, followed by wind turbine controllers to meet the prescribed targets. Each FOWT employs model predictive control to adjust aerodynamic thrust force. The control strategy integrates a dynamic wind farm model that considers floating platform motion and wake transport in changing wind conditions. In a case study with a 1x3 wind farm layout of 5 MW FOWTs, the results show a 25% increase in stable energy production compared to a static layout in one hour for the first scenario. In the second scenario, desired power production was swiftly and consistently achieved.

**Keywords:** Floating offshore wind turbines; Wind farm control; Model predictive control; Dynamic layout optimization.

## 1. Introduction

Offshore wind turbine technologies have gained remarkable global attention because of their ability to tap into rich wind resources, especially in deep-sea environments [1]. In contrast to conventional fixed-bottom offshore structures, which are typically confined to shallower waters due to depth limitations, floating offshore wind turbines (FOWTs) offer the distinct advantage of being deployable in deeper ocean regions [1,2]. This characteristic allows them to harness the potential of more stable and abundant wind energy sources. However, wind turbines tightly clustered within wind farms experience the wake effect which is characterized by reduced wind speeds and high turbulence levels [3]. It was estimated that the wake effects have the potential to decrease power generation from each single downstream turbine by as much as 60%, reducing overall wind farm power production by up to 54% [4–7].

Several advanced real-time control methods have emerged in the technical literature, primarily to address the challenge posed by the wake effects and optimize the overall energy production and structural integrity of the wind farm. The control techniques can be broadly classified into two categories: those requiring hardware modifications [8–10], and those that do not [11–16]. The second control strategy, which eliminates the need for additional hardware, is cost-effective and can be easily implemented in existing utility-scale wind farms.

This strategy directly changes the wind turbine parameters, such as the nacelle yaw angle, generator torque, and collective blade pitch angle. Among the various wind farm control methods explored in existing research which do not require additional hardware, two have received extensive attention: power derating [17–21], often known as axial induction-based control, and yaw-based wake redirection (i.e., wake steering) [6,22–28]. The power derating technique reduces the thrust force of an upstream turbine which creates smaller wake, allowing for improved wind flow conditions and boosting power production in downstream turbines. On the other hand, the yaw-based wake redirection is a wind farm control strategy that involves adjusting the orientation or yaw angle of individual wind turbines to redirect the wake generated by upstream turbines away from downstream turbines, thereby optimizing their performance by reducing the overlapping areas between the wakes and downstream rotors. A promising alternative for enhancing the performance of floating wind farms capitalizes on the added degrees of freedom offered by the floating platform to dynamically optimize the wind farm layout [8,11–16,29–31]. In this approach, a centralized wind farm controller first identifies the optimal wind farm layout to meet the power requirement (i.e., maximization or regulation) and reduce the wake effects [13,14,16,32]. Subsequently, the wind turbine controllers are tasked to achieve the prescribed targets ensuring stable power generation and safe operation [11,12,14–16]. This can be accomplished through two approaches: one involves actively generating the thrust force to reposition the turbine platform based on actuators such as thrusters and winches [8–10,33], necessitating more energy and additional hardware. Alternatively, the second approach involves passive adjustments to the aerodynamic thrust force using existing control inputs, namely the yaw angle, collective blade pitch angle, and generator torque [11–16,29,30]. Compared to other control techniques, the dynamic layout optimization strategy offers significant advantages since it enables the FOWTs to mitigate the wake effects and maximize energy production without the need to reduce the capacities of upstream turbines, as seen in the axial induction-based control, or employ yaw mechanisms to steer wakes which can negatively impact power generation and introduce stability issues. For example, Fleming et al. [31] indicated that the wake deflection through the repositioning technique resulted in a 41% improvement in power generation compared to the yaw and tilt misalignment techniques which yielded a 4.6% and 7.6% increase, respectively. Similarly, other studies have reported significant increase in power generation through layout optimization with improvements of up to 53.5% compared to a non-optimized 3x6 layout [11,12].

Despite the recent efforts to enhance the control technique for repositioning the FOWTs by passively manipulating the aerodynamic thrust force [11–13,16,34], limited attention has been given to the more practical scenario involving time-varying free stream wind velocities. This scenario requires sophisticated models capable of swiftly simulating changing wind conditions and customized control techniques to maintain real-time stability. While various control algorithms, such as proportional-integral-derivative [14], the H∞ state feedback controller [15], and reinforcement learning [35–38], have been applied in wind turbine control, designing robust controllers that can effectively handle multiple-input multiple-output systems with numerous constraints and time-varying environmental disturbances remains a challenging endeavor. In addressing these control challenges, model predictive control (MPC) is often favored over alternative control algorithms due to its ability to optimize control inputs over a future time horizon, taking into account system dynamics and constraints, thus enabling more precise and adaptable control in intricate and dynamic processes. While MPC has been applied to various wind turbine control problems [30,39,40], none of these studies have yet explored its application in the context of a more realistic wind farm scenario featuring time-varying wind conditions.

This study will focus on exploring a dynamic layout optimization strategy aimed at fulfilling power requirements and mitigating wake effects. Two scenarios will be considered: the first prioritizing power maximization and the second focusing on power set-point tracking. In both scenarios, the objective is to minimize the overlap of wakes with downstream rotors. In the proposed methodology, an initial step involves a centralized wind farm controller identifying the optimal wind farm layout to fulfill power requirements (whether maximizing or regulating power) while also mitigating wake effects. This optimization process is accomplished using Matlab/Simulink. Subsequently, the wind turbine controllers are charged with the task of achieving the designated objectives ensuring both stable power generation and safe operation. Each FOWT is equipped with a model predictive control (MPC) which manipulates directly the aerodynamic thrust force using three control inputs, namely the collective blade pitch angle, the generator torque, and the nacelle yaw angle. The proposed control strategy incorporates an efficient dynamic wind farm model [41], which includes the simulation of floating platform motion and wake transport under varying wind conditions and platform movements. In addition, the MPC predictive model is based on a highly efficient dynamic model [42], designed specifically for real-time control applications. A case study of a wind farm consisting of a 1x3 layout is considered for both scenarios (i.e., power maximization and regulation) using a 5 MW offshore semi-submersible baseline wind turbine developed by the National Renewable Energy Laboratory (NREL) in the United States. To evaluate the controller's performance, it will be compared with a wind farm that lacks a repositioning mechanism.

## 2. Model Description

In this section, an overview of the dynamic wind farm model, FOWFSim-Dyn [41], is presented. This model is capable of simulating floating platform motion and the dynamic propagation of wakes under time-varying wind conditions. FOWFSim-Dyn comprises two modules: the wake module and the wind turbine module. The wake module solves a one-dimensional momentum conservation equation to simulate the dynamic propagation of wake centerline locations and average velocities, utilizing a constant temporal wake expansion rate to approximate momentum recovery. On the other hand, the wind turbine module simulates platform dynamics by accounting for aerodynamic, hydrodynamic, and mooring line forces, while computing the power output of the wind farm $P_{farm}(t)$. The block diagram illustrating the interaction between the two modules is presented in Fig. 1. As shown in Fig. 1, the wake module generates the effective wind velocity vector $\mathbf{V}_i(t)$ incident on turbine $i$'s rotor given the states $x(t)$ and inputs $u(t)$ of all turbines, as well as the free stream wind velocity $\mathbf{V}_\infty(t)$ and acceleration vectors $\dot{\mathbf{V}}_\infty(t)$. On the other hand, the wind turbine module calculates $\dot{x}_i(t)$ using the effective wind velocity vector $\mathbf{V}_i(t)$ in conjunction with the turbine states $x(t)$ and inputs $u(t)$.

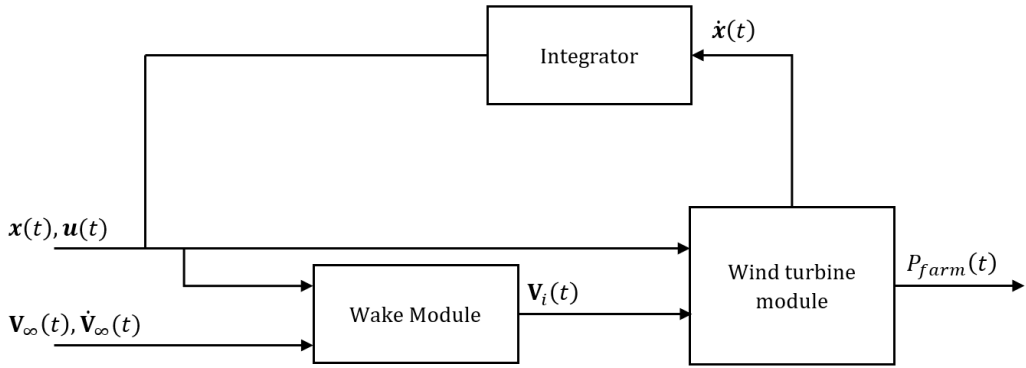

**Fig. 1** Block diagram of the FOWFSim-Dyn model

While the state vector $\boldsymbol{x}(t)$ includes both the position $\mathbf{r}_i(t) \coloneqq [x_i(t) \quad y_i(t)]^T$ and velocity $\mathbf{v}_i(t) \coloneqq [v_{x,i}(t) \quad v_{y,i}(t)]^T$ vectors for all floating wind turbines in the wind farm ($i \in \mathcal{F} = \{1,2,\dots,N\}$ where $N$ represents the total number of floating wind turbines within the wind farm), the input vector $\boldsymbol{u}(t)$ consists of the effective control inputs $\boldsymbol{u}_i(t) = [\beta_i \quad \tau_{g,i} \quad \gamma_i]^T$ for all wind turbines, which are adjusted to meet the control commands. Here, $\beta_i$ represents the collective blade pitch angle, $\tau_{g,i}$ denotes the generator torque and $\gamma_i$ is the nacelle yaw angle.

In this study, a 5 MW offshore three-bladed wind turbine equipped with a semi-submersible platform, has been considered. The selected platform has been developed by the National Renewable Energy Laboratory (NREL) in the United States and has three cylindrical columns linked to mooring lines, along with a central fourth column responsible for supporting the tower. A schematic representation of the wind turbine is given in Fig. 2. Additional information regarding the wind turbine's characteristics is available in references [43] and [44].

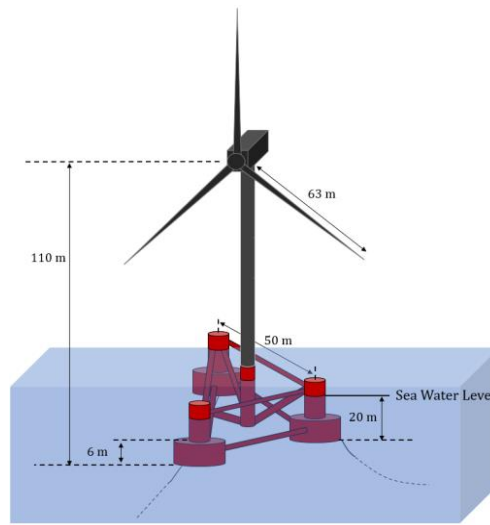

**Fig. 2** Schematic representation of the semi-submersible 5-MW wind turbine

## 2.1 Wake module

The wake module, alternatively referred to as the aerodynamic module, considers both the free stream velocity and the wind farm layout to simulate the evolution of wakes within the wind farm. This simulation yields the effective wind velocities incident on the turbines. Specifically, in a single wake scenario, the wake module's objective is to determine three time-dependent variables: the position of the wake centerline $y_{w,i}$ relative to the $\hat{x}_i$ axis, the average velocity of the wake at the centerline $\mathbf{v}_{w,i}$ and the wake diameter $D_{w,i}$. These three variables of interest are illustrated in Fig. 3.

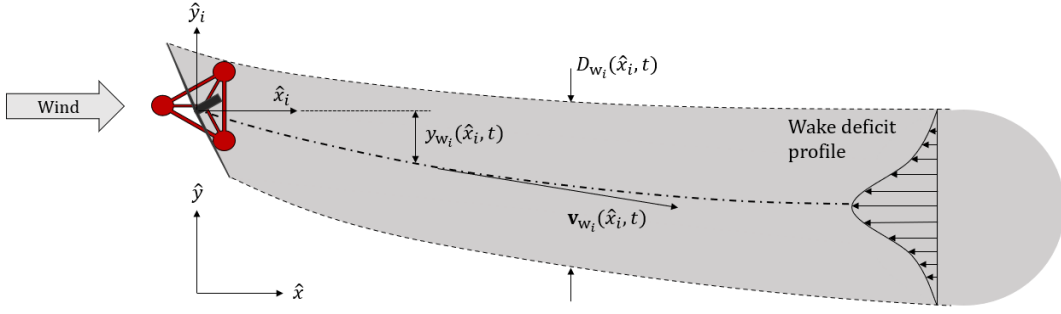

**Fig. 3** Schematic representation of the wake zone generated by the turbine $i$

With a fixed global frame of reference ($\hat{x}$, $\hat{y}$) and a local translating frame ($\hat{x}_i$, $\hat{y}_i$) linked to each wind turbine, the three characteristic variables can be calculated by solving the following equations [41], which are derived from a one-dimensional momentum conservation equation and the simplified assumption of a constant temporal wake expansion rate:

$$\frac{\partial \mathbf{v}_{w,i}(\hat{x}_i,t)}{\partial t} + \left(U_\infty(t) - v_{x,i}(t)\right)\frac{\partial \mathbf{v}_{w,i}(\hat{x}_i,t)}{\partial \hat{x}_i} =$$

$$\dot{\mathbf{V}}_\infty(t) - \dot{\mathbf{v}}_i(t) + \frac{2}{D_{w,i}(\hat{x}_i,t)}\frac{dD_{w,i}(\hat{x}_i,t)}{dt}\left(\mathbf{V}_\infty(t) - \mathbf{v}_i(t) - \mathbf{v}_{w,i}(\hat{x}_i,t)\right) \quad (1)$$

$$\frac{\partial y_{w_i}(\hat{x}_i,t)}{\partial t} + \left(U_\infty(t) - v_{x,i}(t)\right)\frac{\partial y_{w_i}(\hat{x}_i,t)}{\partial \hat{x}_i} = v_{w_i}(\hat{x}_i,t) \quad (2)$$

$$\frac{\partial D_{w_i}(\hat{x}_i,t)}{\partial t} + \left(U_\infty(t) - v_{x,i}(t)\right)\frac{\partial D_{w_i}(\hat{x}_i,t)}{\partial \hat{x}_i} = k_t \quad (3)$$

where $\mathbf{V}_\infty := [U_\infty \quad V_\infty]^T$ represents the free stream wind velocity and $k_t$ is the temporal expansion rate, assumed to be constant. Equations (1), (2) and (3) are subsequently solved using the finite difference method. Once the average velocity of the wake at the centerline, $\mathbf{v}_{w,i}$, has been determined, a Gaussian profile is assumed for the velocity distribution with respect to the radial distance from the centerline of the wake.

To consider the influence of wake interaction in the presence of multiple wind turbines, the effective kinetic energy deficit approach is adopted [45]. As a result, the effective incident wind velocity vector on the rotor of turbine $i$ can be expressed as:

$$\mathbf{V}_i = \left\{\|\mathbf{V}_\infty\| - \sqrt{\sum_{q \in \mathcal{U}_i}\left(\|\mathbf{V}_\infty\| - \bar{\mathbf{v}}_{w,q \to i} \cdot \mathbf{n}_\infty\right)^2}\right\}\mathbf{n}_\infty \quad (4)$$

where $\mathcal{U}_i = \{1, 2, 3, \ldots, i-1\}$ represents the set of indices of wind turbines located upstream of wind turbine $i$, $\bar{\mathbf{v}}_{w,q \to i}$ represents the effective velocity of wake $q$ at wind turbine $i$ which is determined using the gaussian assumption for the wake profile [41,46] and $\mathbf{n}_\infty$ is a unit vector aligned with $\mathbf{V}_\infty$.

### 2.2 Wind turbine module

After obtaining the effective wind speed vector at each wind turbine rotor, the wind turbine module is utilized to simulate the platform dynamics using a Newtonian approach, expressed as follows:

$$\dot{\mathbf{v}}_i(t) = \ddot{\mathbf{r}}_i(t) = \frac{\mathbf{F}_{a,i}(t) + \mathbf{F}_{h,i}(t) + \mathbf{F}_{m,i}(t)}{m_i + m_{a,i}} \quad (5)$$

where $m_i$ represents the wind turbine mass, $m_{a,i}$ is the hydrodynamic added mass associated with the $i^{\text{th}}$ wind turbine, $\mathbf{F}_{a,i}(t)$ is the aerodynamic thrust force, $\mathbf{F}_{h,i}(t)$ is the hydrodynamic force and $\mathbf{F}_{m,i}(t)$ is the mooring line force. The aerodynamic thrust force is directly applied on the turbine's rotor and is given for turbine $i$ as:

$$\mathbf{F}_{a,i}(t) = \frac{1}{8} C_{t,i} \pi \rho D_i^2 \|\mathbf{V}_{\text{rel},i}\|^2 \mathbf{n}_i \tag{6}$$

where $D_i$ is the rotor diameter, $\rho$ is the air density, $C_{t,i}$ is the thrust coefficient, $\mathbf{n}_i$ is a unit vector normal to the rotor and $\mathbf{V}_{rel,i}(t)$ is the incident wind speed experienced by turbine $i$ as $\mathbf{V}_{rel,i}(t) = \mathbf{V}_i(t) - \mathbf{v}_i(t)$. The hydrodynamic force is represented by the sum of drag forces associated with all submerged elements ($j \in \mathcal{D}_i$) of the floating wind turbine, and is expressed as follows:

$$\mathbf{F}_{h,i}(t) = -\frac{1}{2}\left(\sum_{j \in \mathcal{D}_i} C_{d,i,j} A_{d,i,j}\right) \rho_w \|\mathbf{v}_i\| \mathbf{v}_i \tag{7}$$

where $\rho_w$ is the water density while $C_{d,i,j}$ and $A_{d,i,j}$ are respectively the drag coefficient and the submerged area of the $j^{\text{th}}$ component. The restoring force takes into consideration the total number ($\mathcal{M}_i$) of mooring line force vectors acting on the wind turbine $i$ and it is expressed as:

$$\mathbf{F}_{m,i}(t) = \sum_{k \in \mathcal{M}_i} -H_{F,i,k} \frac{\mathbf{r}_{F/A,i,k}}{\|\mathbf{r}_{F/A,i,k}\|} \tag{8}$$

where $H_{F,i,k}$ is the magnitude of the horizontal tension component corresponding to the $k^{\text{th}}$ mooring line of the $i^{\text{th}}$ turbine [41] and $\mathbf{r}_{F/A,i,k}$ is the horizontal position vector from the anchor to the corresponding fairlead of the $k^{\text{th}}$ mooring line.

Using the output from both the wake and wind turbine modules, the total wind farm power can be computed as follows:

$$P_{farm}(t) = \sum_{i \in \mathcal{F}} P_{out,i}(t) \tag{9}$$

where $P_{out,i}$ is the instantaneous power output of turbine $i$ expressed as:

$$P_{out,i}(t) = \tau_{g,i} \omega_{g,i} \eta_{g,i} \tag{10}$$

where $\omega_{g,i}$ is the generator speed of turbine $i$ and $\eta_{g,i}$ is the generator conversion efficiency of turbine $i$.

## 3. Control Design

### 3.1 Control structure

To attain the targeted power production and mitigate the wake effects, the proposed controller system is structured into two primary tiers: the wind farm controller and the wind turbine controller, as depicted in Fig. (4). Further insights into each controller are elaborated upon in the subsequent sections.

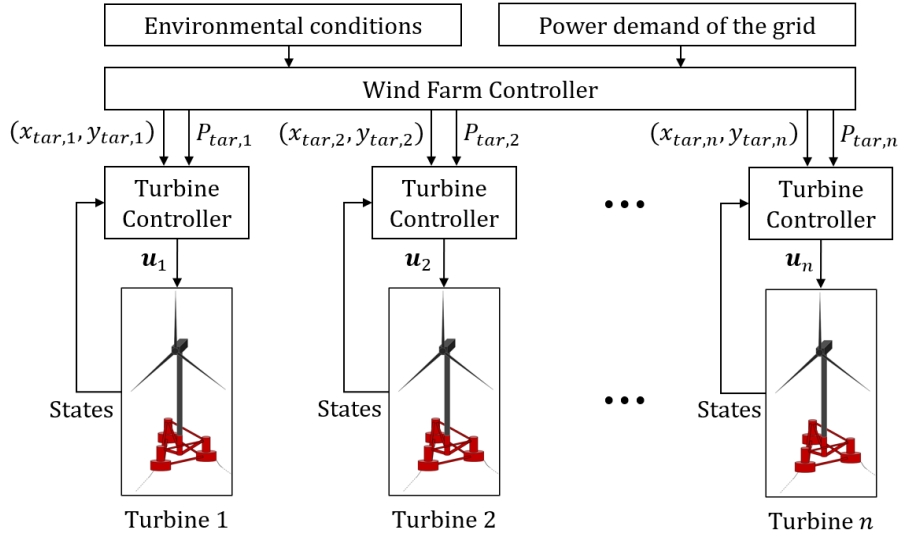

**Fig. 4** Block diagram of the control system

### 3.1 Wind farm controller

The wind farm controller calculates the optimal setpoints (i.e., optimal wind turbine coordinates and individual power target), considering environmental conditions (such as wind) and the power demand from the electrical grid. Achieving these optimal setpoints requires solving an optimization problem. For instance, in the context of power maximization, the optimization problem can be formulated as follows:

$$\max_{setpoints} (P_{farm}); \text{ subject to } \begin{cases} \text{safety limits} \\ \text{system contraints} \\ \text{wake limitation} \\ \dots \end{cases} \quad (11)$$

Several constraints must be met to ensure the reliable performance and safe operation of the floating wind turbine in the presence of environmental disturbances. Furthermore, to enhance wind turbine performance and prolong its operational lifespan, it is crucial to guide the controller to minimize, whenever feasible, the areas affected by wakes, which are known to be a source of various fatigue-related issues. In this study, the optimization problems are resolved using the Simulink Design Optimization toolbox [47], employing the pattern search algorithm [48]. Specifically, this iterative heuristic method relies on successive simulations (i.e., episodes) to explore the search space. The objective function is then evaluated at the end of each episode to obtain the optimal solutions, which will subsequently be employed as setpoints in the various control scenarios. In this study, the duration of each of these episodes are fixed at 2000 seconds to allow for a complete repositioning, while the collective blade pitch angle $\beta_i$ and the generator torque $\tau_{g,i}$ are replaced by a constant axial induction factor $a = 0.3$, obtained as an average value during turbine operation, for each wind turbine. This substitution is made to avoid dealing with several control inputs during the acquisition of optimal solutions and, consequently, to facilitate the resolution of the optimization problems.

### 3.2 Wind turbine controller

The wind turbine controller associated with the $i^{\text{th}}$ turbine is designed to reposition the floating platform, aligning it with its designated coordinates $(x_{tar,i}, y_{tar,i})$, and to maintain the generated power around the specified target value $(P_{tar,i})$, as directed by the wind farm controller (Fig. 4). In this study, the control strategy requires no external actuators or additional hardware, as it directly manipulates the aerodynamic thrust force. This feature makes it suitable for both new and existing offshore wind turbines. Each wind turbine $i$ has three control inputs, namely the collective blade pitch angle $\beta_i$, the generator torque $\tau_{g,i}$, and the nacelle yaw angle $\gamma_i$. The collective blade pitch angle adjusts the magnitude of the aerodynamic force by controlling the thrust coefficient and the tip-speed ratio. The generator torque regulates the generator speed and power extraction. On the other hand, the nacelle yaw angle modifies the rotor's orientation, thereby adjusting the direction of the aerodynamic thrust force.

The wind turbine controller comprises two primary subsystems: the power regulator, responsible for maintaining the specified power level $P_{tar,i}$ consistently, and the position controller, which governs the platform's movement, as depicted in Fig. 5. Further details regarding each control subsystem will be presented in the following sections.

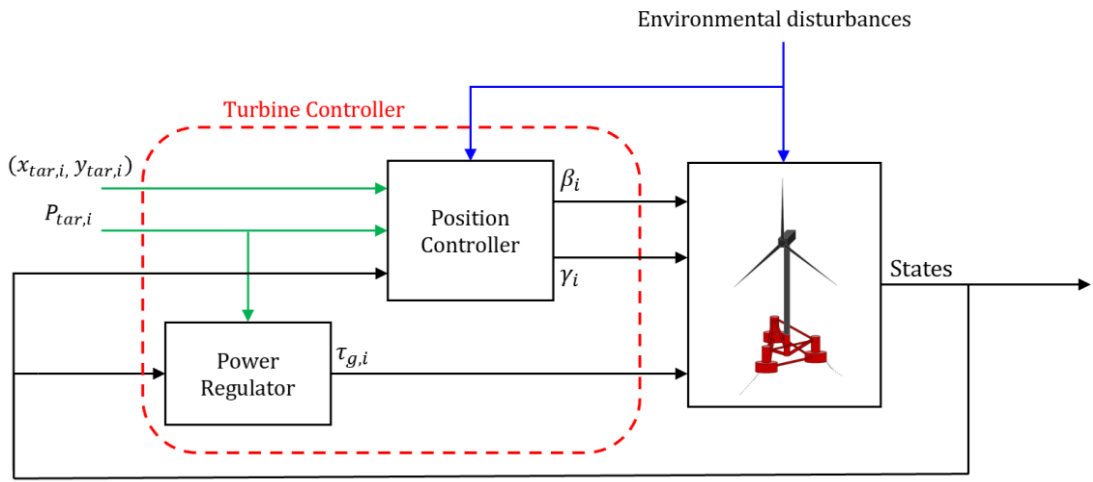

**Fig. 5** Block diagram of the wind turbine controller

### 3.2.1 Power regulator

The control law governing the power regulator is based on the constant power strategy [43]. Specifically, for the wind turbine $i$, the regulator provides the torque $\tau_{g,i}$ necessary to generate the desired power level $P_{tar,i}$ as determined by the following equation:

$$\tau_{g,i} = \frac{P_{tar,i}}{\eta_{g,i}\omega_{g,i}} \tag{12}$$

where $\omega_{g,i}$ is the generator speed and $\eta_{g,i}$ is the generator's conversion efficiency.

### 3.2.2 Position controller

The goal of the position controller is to relocate the floating platform of turbine $i$ to the target position $(x_{tar,i}, y_{tar,i})$, as provided by the wind farm controller. To achieve this objective, the controller receives the current system states, such as position, along with perturbations and generates appropriate control inputs $\beta_i$ and $\gamma_i$ to modify the aerodynamic thrust force. In this study, a model predictive control (MPC) is employed to achieve these control objectives. The MPC model relies on a dynamic representation of the system's process. This internal representation is used to aid in predicting the necessary control actions while considering the system's constraints.

In this study, an efficient control-oriented dynamic model [42] suitable for real-time control applications has been employed as the prediction model for the MPC controller. This selection was made to minimize the computational costs and facilitate controller design. The model represents the overall dynamics of the floating wind turbine while reducing the number of degrees of freedom (DOFs) to six platform DOFs and two drivetrain DOFs, resulting in a total of 15 states. The derivation of this model is based on a Newtonian approach, assuming that the floating wind turbine can be treated as a rigid body subjected to various forces, including aerodynamic, hydrodynamic, buoyancy, and mooring line forces, along with their respective torques, as illustrated in Fig. 6. The model is expressed in a state-space format as follows:

$$\dot{X} = f(X, u, v, w) \tag{13}$$

where $u$ is the control inputs vector, $v$ and $w$ are the vectors referring to the environmental disturbances, with $v$ corresponding to the wind and $w$ to the waves. Finally, the vector $X$ refers to the system states, which include translational coordinates of the floating platform (surge $x$, sway $y$ and heave $z$), rotational coordinates (roll $\theta_x$, pitch $\theta_y$ and yaw $\theta_z$), their derivatives, rotor rotational speed $\omega_r$, generator rotational speed $\omega_g$ and the shaft deflection angle $\Delta\theta_r$. The updated system states can be determined through straightforward integration of $\dot{X}$, derived from Eq. (13). Further details regarding the model and its parameters can be retrieved from [42].

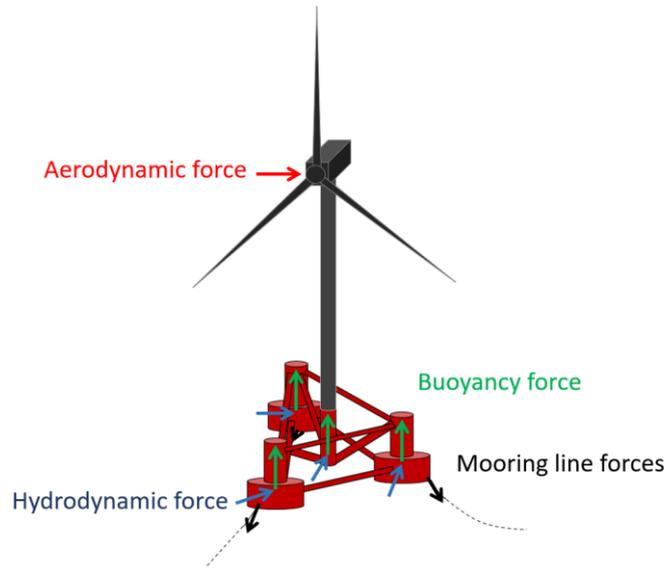

**Fig. 6** Schematic representation of the forces acting on a FOWT

The nonlinear model represented by Eq. (13) is linearized around carefully chosen equilibrium points, denoted as $p_{eq} = [X_0, u_0, v_0, w_0]^T$ by solving the following equation:

$$f(X_0, u_0, v_0, w_0) = 0 \tag{14}$$

This results in the subsequent linearized model, represented as:

$$\delta\dot{X} = A\delta X + B\delta u_{MPC} + C\delta v \tag{15}$$

where $u_{MPC} = [\beta \quad \gamma]^T$ represents the MPC control inputs while $A$, $B$ and $C$ are the equivalent linearized matrices which are determined based on the Jacobians of the function $f$ at the equilibrium point $p_{eq}$ as:

$$A = \left.\frac{\partial f}{\partial X}\right|_{p_{eq}} \quad B = \left.\frac{\partial f}{\partial u_{MPC}}\right|_{p_{eq}} \quad C = \left.\frac{\partial f}{\partial v}\right|_{p_{eq}} \tag{16}$$

Based on the linearized model, the MPC controller employs an iterative approach to determine the optimal control actions. Specifically, the controller utilizes a discrete-time approach, seeking the optimal sequence of control inputs over the control horizon $N_m$ that minimizes a cost function $J$ over the prediction horizon $N_p$ while accounting for the system constraints. This process is then repeated at each time step. The cost function $J$ can be expressed as:

$$J = \sum_{k=i+1}^{i+N_p} \left\| Q[Y_k - Y_{ref}] \right\|^2 + \sum_{k=i}^{i+N_m-1} \left\| R[u_{MPC,k} - u_{MPC,k-1}] \right\|^2 + \sum_{k=i}^{i+N_p-1} \left\| S[u_{MPC,k} - u_{MPC_{ref,k}}] \right\|^2 \quad (17)$$

where $Q, R$ and $S$ are weight matrices which are designed to penalize excessive outputs errors and regulate the excessive use of the control inputs (i.e., blade pitch and nacelle yaw), $Y$ is the output vector, representing the desired states being controlled (e.g., surge, sway and generator rotational speed), which is optimized to match the target vector $Y_{ref}$ and $u_{MPC_{ref,k}}$ are the reference control inputs which are obtained based on the equilibrium point corresponding to the target position. It's important to highlight that the control law at each time step $k$ is subject to various constraints to ensure compliance with saturation and rate limits for control inputs. Additionally, these constraints prevent the system from surpassing the prescribed minimum and maximum permissible displacements and generator rotational speeds [30,43].

## 3. Case Study

To demonstrate the performance of the proposed control strategy for floating wind turbines under varying wind conditions, two distinct case scenarios will be analyzed using the dynamic wind farm model (FOWFSim-Dyn) and Matlab/Simulink. In the first scenario, the focus is on maximizing power production, whereas the second scenario involves a specified target power output. Both scenarios involve a 1x3 wind farm layout with 5 MW FOWTs, initially aligned in the wind direction ($\hat{x}$) and separated by a distance equivalent to seven times the diameter of their rotor, as depicted in Fig. 7.

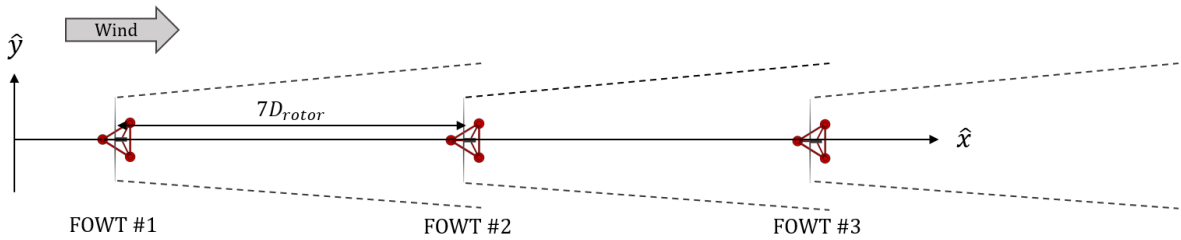

**Fig. 7** Floating wind farm configuration for the control scenarios

Each FOWT is equipped with an MPC controller, which receives instructions from the centralized wind farm controller. To extend the movable range of the platform, the mooring cables for each floating platform have been adjusted to a length of 920 meters, compared to the initial length of 835.5 meters [12,13,29]. It should be noted that the longitudinal platform position ($x_i$) is not actively controlled in this application due to its minimal variation [12]. Therefore, it does not affect significantly the power production, unlike the lateral platform position ($y_i$). However, constraints are still applied to the longitudinal platform position $x_i$ to ensure a stable and safe operation of the system.

Furthermore, while the FOWFSim-Dyn model has the capability to accommodate fluctuating wind conditions over time, it only considers a calm sea state without waves as the basis of its operation [41], therefore the waves were not considered in this study. As a result, the linearized model of the MPC controller was developed solely based on wind speed. Nevertheless, it's worth noting that the suggested controller has the capability to incorporate both wind and wave disturbances with minimal adjustments to the control parameters. In both scenarios, the wind conditions were determined for a one-hour duration using the Von Karman turbulence spectrum, where the mean incident wind speed at the rotor hub height was set at $\bar{V}_\infty = [14, 0]$ [m/s], with a turbulent intensity of 0.036 and an integral length scale of 170 [m]. It is important to emphasize that when wind speeds are excessively low, the repositioning mechanism, which relies on directly manipulating the aerodynamic thrust force, may not be effective. In fact, during periods of low wind speeds, the thrust force might be too weak to compensate the effects of other forces, such as those from the mooring lines. Consequently, specific repositioning tasks related to power demand may not be achievable. To address these challenges, potential solutions include increasing the length of the mooring lines to reduce the magnitude of the restoring force or incorporating additional actuators to aid in repositioning the wind turbine under such conditions [13,15,16,29].

To maintain the wind turbine within its designated operating range [43], a series of constraints must be enforced. This involves imposing restrictions on the generator's rotational speed, $\omega_g$, to ensure it falls within the range of 669.3 [rpm] to 1173.7 [rpm]. Additionally, saturation and rate-limits on control inputs, as outlined in Table (1), are enforced. It's important to highlight that a maximum value for the collective blade pitch angle, $\beta_i$, is set at 0 degrees. This decision is made to prevent any value above 0 degrees, which could potentially occur at lower wind speeds. Exceeding this value would trigger a shift in the control logic from pitch-to-stall regulation to pitch-to-feather regulation. This nonlinear transition can lead to instability in the operation of the wind turbine and result in inappropriate control responses as the controller relies on a linear prediction model.

**Table 1.** Imposed constraints, saturations, and rate limits on each controller in the control process

| Control Inputs | Saturation | Rate Limit |
| --- | --- | --- |
| $\beta_i$ | [−30, 0] [deg] | [−8, 8] [deg/s] |
| $\tau_{g,i}$ | [0, 47.402] [kN·m] | [−15, 15] [kN·m/s] |
| $\gamma_i$ | [−60, 60] [deg] | [−0.3, 0.3] [deg/s] |

The predictive models associated with the MPC controllers were linearized at multiple equilibrium points, including both the initial and target positions. These positions are determined by the centralized wind farm controller based on the selected scenario. Subsequently, the wind turbine controller will transition from one equilibrium point to another in accordance with the designated target position. The MPC parameters, such as prediction and control horizons as well as weight matrices, were determined through a trial-and-error process. Initially, the FOWTs are oriented along the $\hat{x}$-axis with starting positions of $(x_1, x_2, x_3) = (84, 966, 1848)[m]$ and $(y_1, y, y_3) = (0, 0, 0)[m]$. The mean power output is $(P_{out,1}, P_{out,2}, P_{out,3}) = (5, 3, 3)[MW]$, clearly indicating the substantial impact of wake effects, particularly on the downstream wind turbines.

### 3.1 First scenario

In this scenario, the primary objective is to maximize the total power output from the wind farm, denoted as $P_{farm}$ (Eq. 9). This goal must be achieved while adhering to system constraints and mitigating the impact of wake effects. Consequently, the centralized wind farm controller's task is to determine the optimal positions for each wind turbine, specifically their lateral positions defined by their respective yaw angles. Subsequently, the wind turbine controllers are tasked to achieve the prescribed targets. An alternative approach to formulating the optimization problem is to instruct the centralized controller to determine optimal turbine locations by maximizing the sum of the effective wind velocity projected onto the rotor-swept area, which is given for wind turbine $i$ as $V_{\perp,i} = \|\mathbf{V}_{\text{rel},i}\| \cos(\gamma_i - \theta_{rel,i})$ where $\theta_{rel,i}$ represents the relative wind angle of $\mathbf{V}_{\text{rel},i}$ with respect to the $\hat{x}$-axis. Details of the solution to this optimization problem and the performance of the wind turbine control system are provided in Sect. 3.1.1, 3.1.2 and 3.1.3.

### 3.1.1 Position control results

The resulting target lateral positions for the 1st, 2nd and 3rd wind turbines were 72, -89, and 70 meters, respectively. The time series of the lateral positions (i.e., sways), using the individual MPC controllers, are illustrated in Fig. 8. It can be concluded that the MPC controllers have effectively and promptly adjusted the lateral positions of the wind turbines to align with their designated target values. Specifically, the target positions were attained within 580 [s] for the 1st, 830 [s] for the 2nd, and 835 [s] for the 3rd wind turbine. These positions were then consistently maintained until the end of the scenario. The obtained root mean square error (RMSE) with respect to the target positions for the lateral displacement were 1.82 [m], 1.97 [m], and 0.85 [m], for the 1st, 2nd and 3rd wind turbines respectively. These results underline the controllers' remarkable accuracy in achieving the desired positions. It's worth noting that a deliberate delay of 500 seconds (determined through trial and error) was introduced for the 3rd turbine to ensure smooth repositioning and prevent potential issues during periods of low wind speeds.

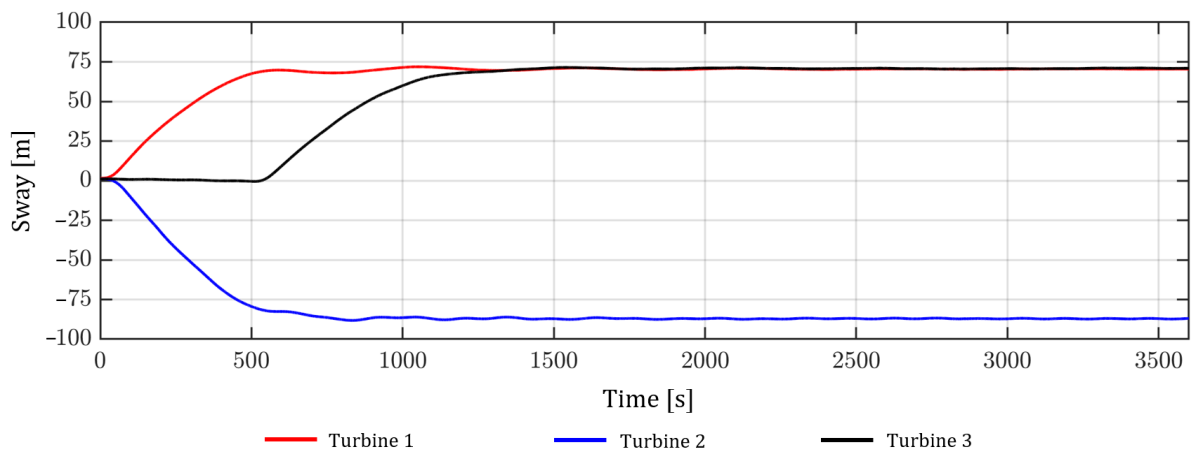

**Fig. 8** Crosswind platform displacements under the 1st control scenario

### 3.1.2 Power control results

To evaluate the performance of the wind farm controller, it will be compared to a wind farm without a repositioning mechanism, where the turbines are kept at their initial starting positions. The time series of the effective wind speeds experienced by each turbine (i.e., $V_{\perp_i}$) are plotted in Fig. 9 for both scenarios.

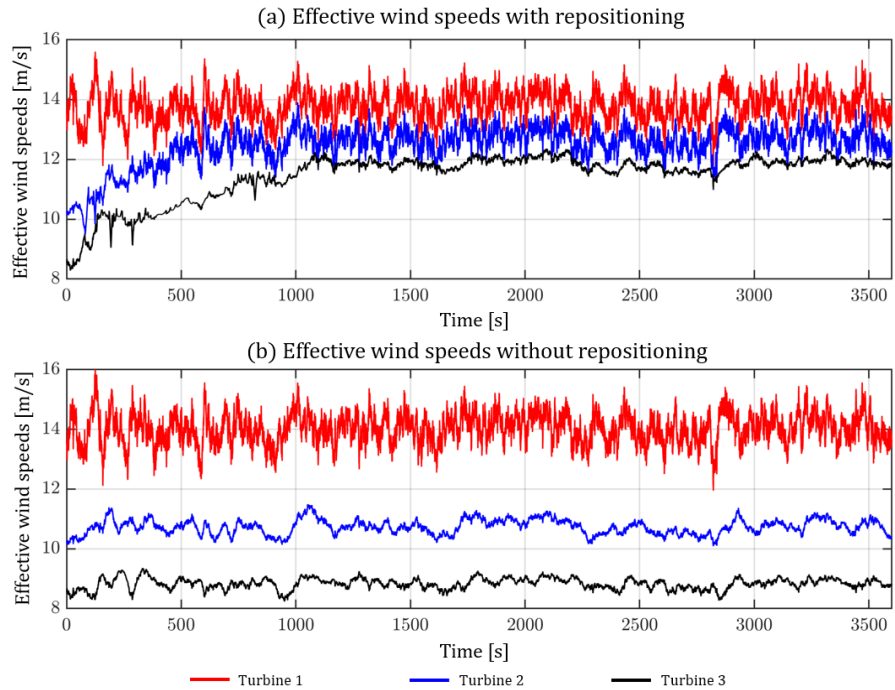

**Fig. 9** Time series of the effective wind speeds experienced by each FOWT for the 1st control scenario under two conditions: (a) with repositioning and (b) without repositioning

It can be concluded that the wind turbine repositioning mechanism plays a substantial role in mitigating wake effects, as the effective wind speeds perceived by downstream turbines are significantly higher compared to a scenario without repositioning. Specifically, after the repositioning of each wind turbine (starting from 1335 seconds), the average effective wind speeds perceived by the 2nd and 3rd downstream turbines have increased by 17.88% and 34.16%, respectively. Furthermore, there is a slight reduction in the effective wind speed perceived by the upstream turbine, which can be attributed to the nacelle misalignment, necessary for platform repositioning. While this misalignment does marginally decrease the perceived wind speed, the impact on power production remains negligible.

To highlight the significant impact of the wake effects, the velocity contour plots at the end of the platform repositioning (i.e., at time $t = 3500$ [s]) are plotted in Fig. 10. It can be concluded that, unlike the scenario without repositioning, the proposed control strategy has successfully relocated the platforms to avoid the wake regions and hence increase the perceived wind speeds which will ultimately maximize the power generation. By avoiding the wake regions, which are characterized by reduced wind speeds and significant fatigue loads, both the power production and service life can be increased.

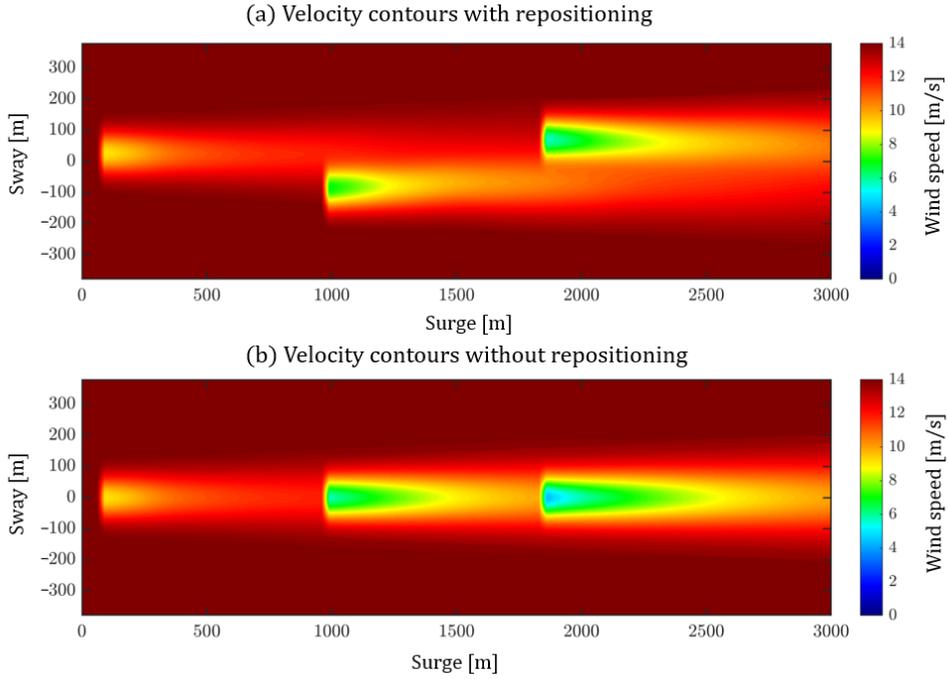

**Fig. 10** Velocity contour plots at time step $t = 3500$ [s] for the 1$^{st}$ control scenario under two conditions: (a) with repositioning and (b) without repositioning

Similar to the perceived FOWT wind speeds, the total power output of the wind farm has also reached its peak value (∼15 MW) as depicted in Fig. 11. Specifically, the repositioning of the platforms has led to a remarkable 25.05% increase in energy production over the course of one hour compared to a scenario where repositioning was not employed. This substantial gain outperforms many existing control-based strategies aimed at mitigating wake effects [17,31].

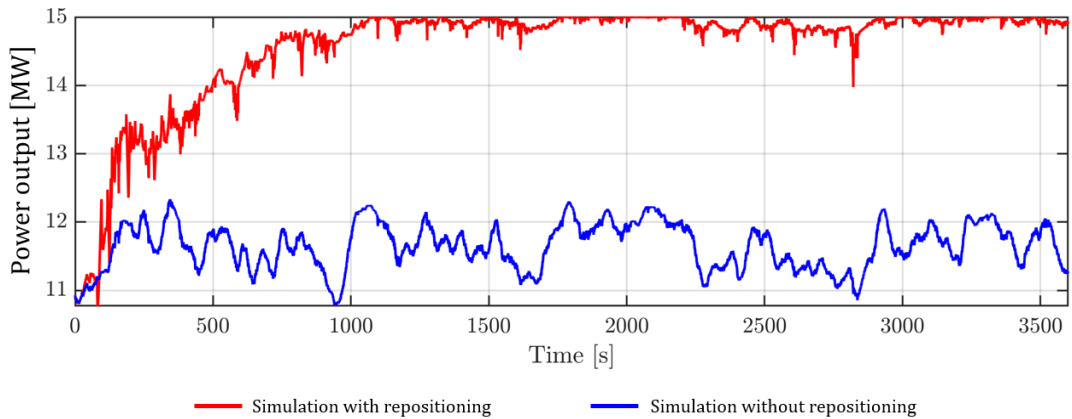

**Fig. 11** Time series of the wind farm total power output for the 1$^{st}$ control scenario under two conditions: with repositioning (red) and without repositioning (blue)

### 3.1.3 Control inputs results

The time series representations of the control inputs for the three MPC controllers are presented in Fig. 12. It can be concluded that all control inputs have remained within the prescribed safety limits while adhering to the constraints detailed in Table 1. In addition, the nacelle yaw angles exhibit a comparable trend to that of the crosswind platform displacements. The MPC controllers have swiftly adapted all control inputs to effectively address the control scenario. Clearly, the 3$^{rd}$ wind turbine required additional time to adapt its control inputs due to the imposed delay (i.e., 500 s). Moreover, owing to the initially low wind speeds experienced by the downstream turbines during the early stages of the simulation, the collective blade pitch angles approached values close to zero degrees. This adjustment was made to prevent stall and optimize wind energy extraction under these specific conditions. As previously mentioned, a strict upper limit of 0 degrees has been imposed on the collective blade pitch angle. Exceeding this threshold triggers a transition in the control logic from pitch-to-stall regulation to pitch-to-feather regulation that cannot be properly managed by the controller. As the perceived wind speeds for the downstream turbines increase, the collective blade pitch angles stabilize at values below zero degrees. This adjustment is made to prevent excessive kinetic energy extraction from the wind through aerodynamic stall. It also facilitates the platform repositioning process while aligning with the power production target.

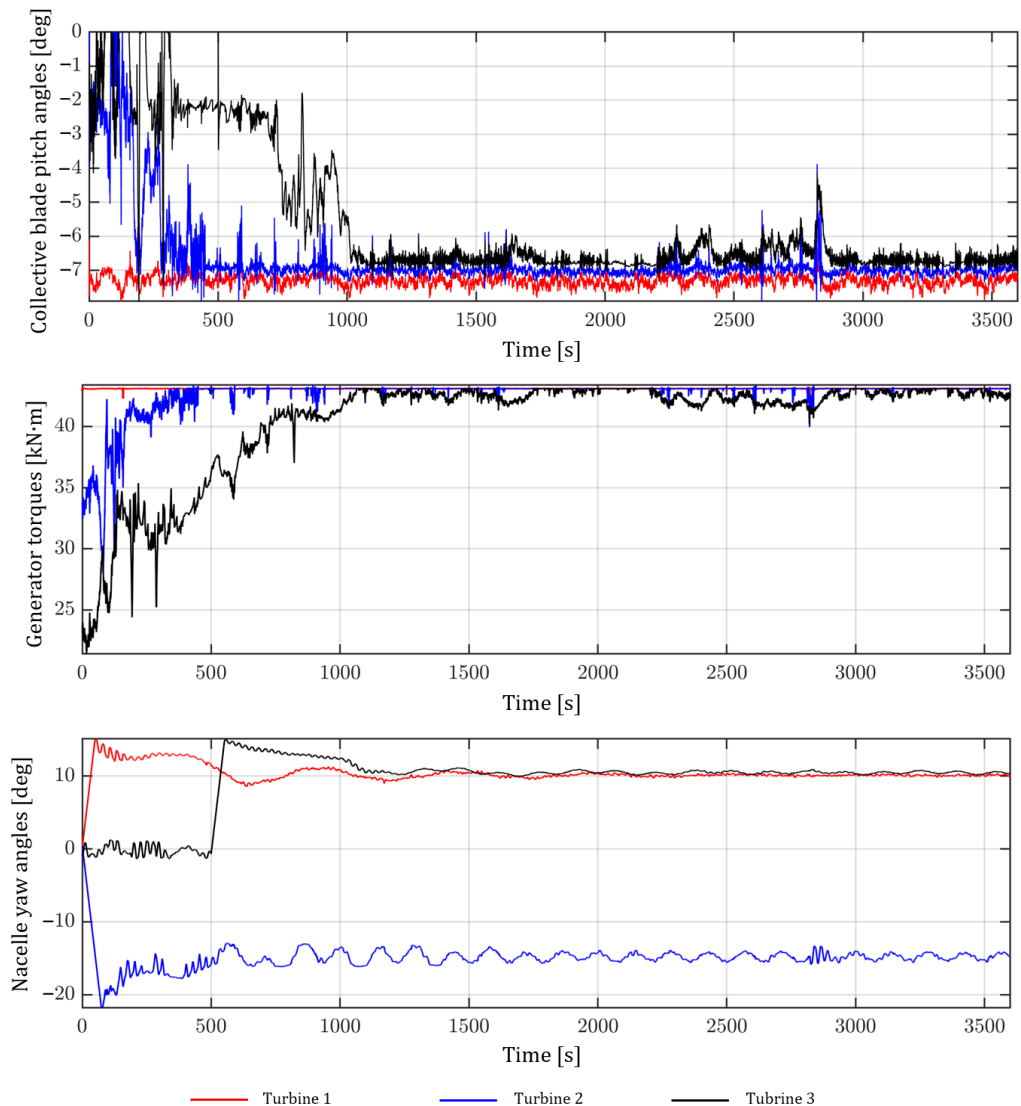

**Fig. 12** Time series of the control inputs for the 1st control scenario

### 3.2 Second scenario

In this particular scenario, the wind farm is required to achieve a target total power output of $P_{farm,tar} = 13$ [MW], representing a realistic power demand from the electrical grid. Additionally, the control strategy must operate within the system constraints while addressing wake effects mitigation. Furthermore, the optimization problem will prioritize minimizing yaw angles to discourage substantial displacements while evading wake regions. All other constraints and system parameters remain consistent with the previous scenario. It's worth noting that, unlike the first scenario, the imposed delay for the 3$^{rd}$ wind turbine has been eliminated. In the subsequent sections, only the results related to position, velocity contours, and power will be presented, as the overall system behavior remains largely consistent with the previous case.

### 3.2.1 Position control results

The target lateral positions achieved were 45, -82, and 45 [m] for the 1$^{st}$, 2$^{nd}$ and 3$^{rd}$ wind turbines, respectively. Figure 13 illustrates the time series of these lateral positions. In this scenario as well, each wind turbine platform successfully reached and maintained its designated position target as instructed by the centralized wind farm controller. These position targets were attained after 630 [s], 840 [s], and 820 [s], respectively, and remained stable until the end of the simulation. The RMSE values for the 1$^{st}$, 2$^{nd}$ and 3$^{rd}$ wind turbine platforms were 0.85 [m], 0.37 [m], and 0.64 [m], respectively. It's worth mentioning that the relatively observed lower errors can be attributed to the controllers finding it comparatively easier to reach and maintain these positions. This ease is primarily because these positions are in close proximity to the wind turbines' initial locations, requiring minimal rotor misalignment. In addition, no delay was required for this scenario, as the 3$^{rd}$ wind turbine was able to achieve and maintain its final position without any observed operational instability.

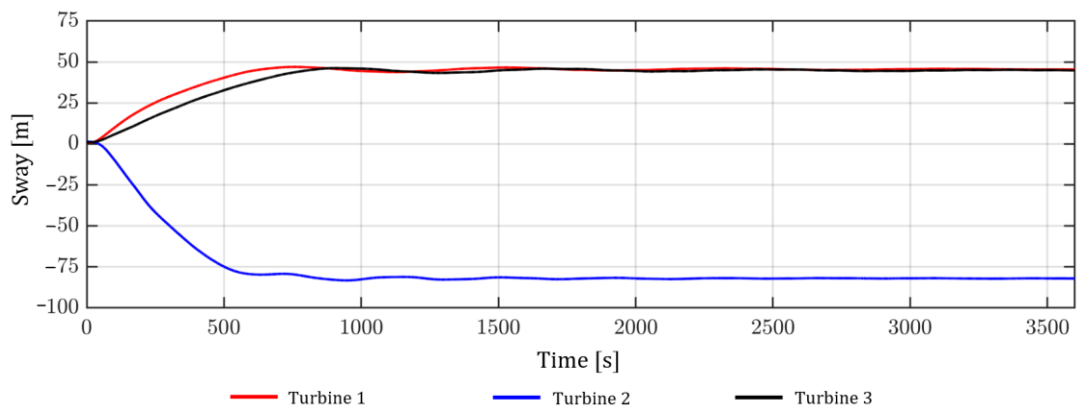

**Fig. 13** Crosswind platform displacements under the 2$^{nd}$ control scenario

### 3.2.2 Power control results

Just as in the 1$^{st}$ scenario, velocity contour plots are provided in Fig. 14 at the end of the platform repositioning (i.e., at time $t = 3500$ [s]) for both the repositioning and non-repositioning cases. It can be concluded that the control strategy has successfully relocated the platforms, avoiding wake-prone regions to alleviate fatigue loads while satisfying the control objectives and constraints.

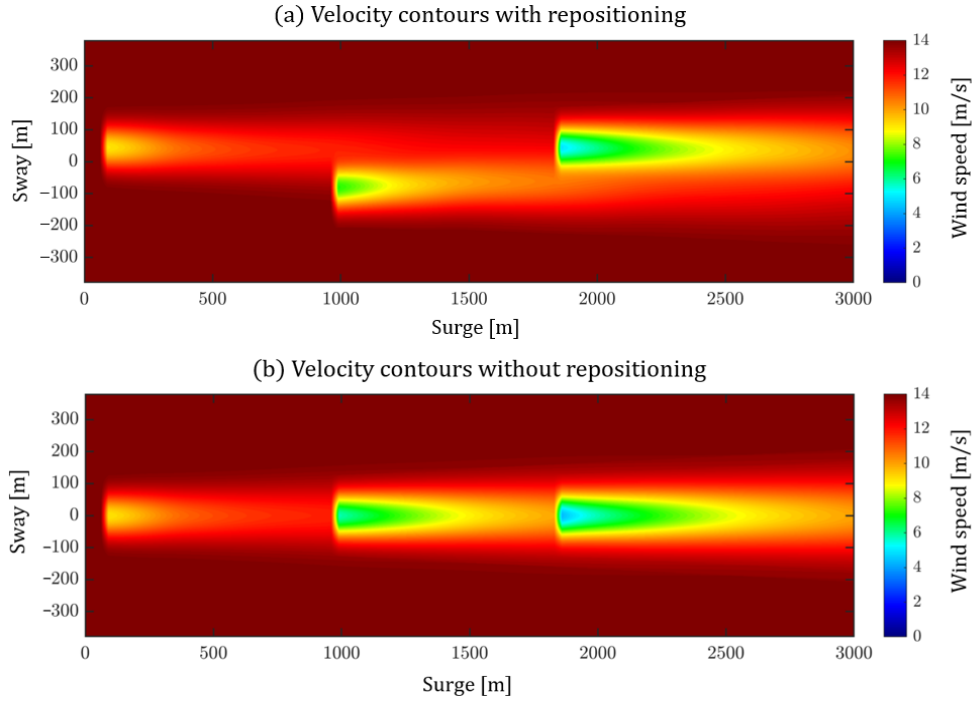

**Fig. 14** Velocity contour plots at time step $t = 3500$ [s] for the 2$^{nd}$ control scenario under two conditions: (a) with repositioning and (b) without repositioning

The wind farm's total power production has successfully achieved its specified target of ~13 MW, as shown by the power generation time series depicted in Fig. 15. In contrast, the scenario without repositioning failed to reach the 13 MW target at any point. Furthermore, the power generation in the repositioning scenario demonstrated increased stability, enhancing its suitability and reliability for meeting the requested power demand from the electrical grid.

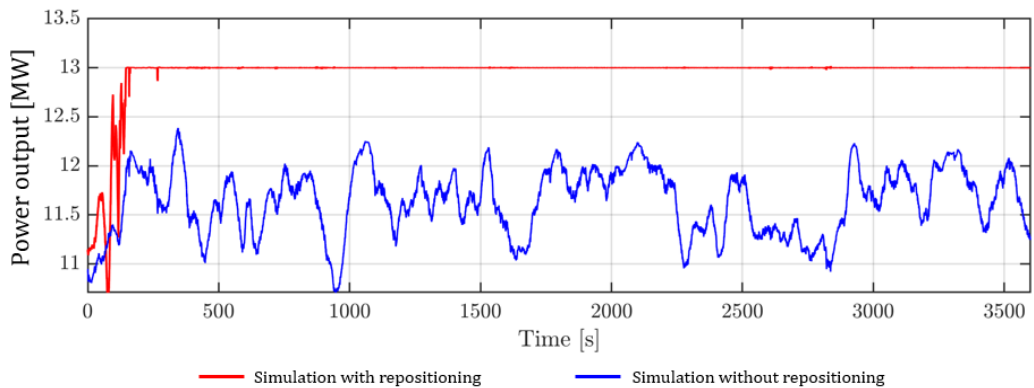

**Fig. 15** Time series of the wind farm total power output for the 2$^{nd}$ control scenario under two conditions: with repositioning (red) and without repositioning (blue)

**Conclusion**

In this study, the dynamic layout optimization approach for offshore floating wind farms was investigated to meet power requirements and alleviate wake effects. Two scenarios were considered: one that prioritized power maximization and another that focused on power set-point tracking. The initial step of the proposed control methodology involved a centralized wind farm controller tasked with identifying the optimal wind farm layout to fulfill power requirements while effectively mitigating wake effects. This optimization process was executed using Matlab/Simulink. Subsequently, wind turbine controllers were asked to achieve the specified objectives. Each Floating Offshore Wind Turbine (FOWT) was equipped with a Model Predictive Control (MPC) system, which directly manipulated the aerodynamic thrust force through three control inputs: collective blade pitch angle, generator torque, and nacelle yaw angle. The proposed control strategy integrated an efficient dynamic wind farm model, which simulates the floating platform motion and wake transport under varying wind conditions. Furthermore, the MPC predictive model was based on a highly efficient dynamic model specifically designed for real-time control applications. To highlight the effectiveness of the proposed approach, a case study was conducted using a 1x3 layout wind farm configuration. Both power maximization and regulation scenarios were evaluated using 5 MW offshore semi-submersible baseline wind turbines. The obtained results demonstrate the successful repositioning of wind turbines in both scenarios with low root mean square errors, effectively avoiding wake regions. This achievement resulted in a remarkable 25% enhancement in stable energy production when compared to a static layout configuration within a one-hour timeframe for the first control scenario. Moreover, in the second scenario, the intended power production was promptly achieved and consistently sustained throughout the control scenario. Due to its reliable control performance and rapid response time, the proposed control approach can be conveniently integrated into both new and existing floating wind farms without requiring any extra hardware.

**Acknowledgements**

The authors are grateful for the financial support provided by the Natural Sciences and Engineering Research Council of Canada (NSERC).